\documentclass[review,sort&compress]{elsarticle}
\usepackage{lineno,hyperref,amsmath,xspace,graphicx}
\usepackage[left=2cm,right=2cm]{geometry}

\modulolinenumbers[5]

\journal{Reviews in Physics}

\begin{document}

\newcommand{\pt}{\ensuremath{p_{\mathrm{T}}}\xspace}
\newcommand{\PTm}{\ensuremath{{p}_\mathrm{T}\hspace{-1.02em}/\kern 0.5em}\xspace}
\newcommand{\HT}{\ensuremath{H_{\mathrm{T}}}\xspace}

\begin{frontmatter}

\title{The experimental status of direct searches for exotic physics beyond the standard model at the Large Hadron Collider}
\author[mymainaddress]{Salvatore Rappoccio}
\ead{Salvatore.Rappoccio@cern.ch}

\address[mymainaddress]{University at Buffalo, State University of New York, \\239 Fronczak Hall, Amherst, NY, USA 14260}

\begin{abstract}
The standard model of particle physics is an extremely successful
theory of fundamental interactions, but it has many known
limitations. It is therefore widely believed to be an effective field
theory that describes interactions near the TeV scale. A plethora of
strategies exist to extend the standard model, many of which contain
predictions of new particles or dynamics that could manifest in
proton-proton collisions at the Large Hadron Collider (LHC). As of
now, none have been observed, and much of the available phase space
for natural solutions to outstanding problems is excluded. If new
physics exists, it is therefore either heavy (i.e. above the
reach of current searches) or hidden (i.e. currently indistinguishable
from standard model backgrounds). We summarize the existing searches,
and discuss future directions at the LHC.
\end{abstract}


\end{frontmatter}


\section{Introduction}

\begin{quotation}
{\it A man said to the universe: 

``Sir, I exist!''

``However,'' replied the universe, 

``The fact has not created in me 

A sense of obligation.''}

\hspace*{6mm} -- Stephen Crane
\end{quotation}

Particle physics is at a crossroads. 
The standard model (SM) explains a wide range of phenomena spanning
interactions over many orders of magnitude, yet no
demonstrated explanation exists for a variety of fundamental
questions.   
Most recently, the discovery of the Higgs
boson~\cite{Englert:1964et,Higgs:1964ia,Higgs:1964pj,Guralnik:1964eu,Higgs:1966ev,Kibble:1967sv,Aad:2012tfa,Chatrchyan:2012xdj,Aad:2015zhl}
at the ATLAS~\cite{Aad:2008zzm} and CMS~\cite{Chatrchyan:2008zzk} detectors has addressed the
mechanism of electroweak symmetry breaking, but
there is no explanation for why the scale of its mass is so much different from naive
quantum-mechanical expectations (the ``hierarchy
problem'')~\cite{Martin:1997ns,Lykken:1996xt,Lane:2002wv,Hill:2002ap,ArkaniHamed:1998rs,Randall:1999ee,Randall:1999vf,ArkaniHamed:2001nc,Strassler:2006im}. Dark
matter (DM) remains an
enigma, despite extensive astronomical confirmation of its
existence~\cite{Bertone:2004pz,Feng:2010gw,Porter:2011nv}. Neutrino
masses are observed to be
nonzero~\cite{Ahmad:2001an,Fukuda:1998mi,Ahn:2012nd,Abe:2013xua}, 
and elements of the Pontecorvo--Maki--Nakagawa--Sakata matrix~\cite{Pontecorvo:1957qd,Maki:1962mu} have been measured,
but these masses are not easily accounted for in the SM~\cite{Capozzi:2016rtj}. 
Unification of the strong and electroweak forces is expected, but not
yet observed nor
understood~\cite{Pati:1973uk,Pati:1974yy,Georgi:1974sy,Murayama:1991ah,Fritzsch:1974nn,Senjanovic:1982ex,Frampton:1989fu,Frampton:1990hz,Schrempp:1984nj,Dimopoulos:1979es,Dimopoulos:1979sp,Buchmuller:1986zs,Eichten:1979ah,Hewett:1988xc};
such models often predict the existence of yet-to-be-observed
leptoquarks (LQs) or proton decay~\cite{Nath:2006ut}. 
Furthermore, there are unexpected observations that are not explained in
the SM, such as the baryon
asymmetry~\cite{Canetti:2012zc}, anomalies in the
decays of bottom-quark hadrons~\cite{Buttazzo:2017ixm}, a discrepancy in
the anomalous magnetic moment of the muon (g-2)~\cite{Holzbauer:2016cnd}, and the
strong CP problem~\cite{Peccei:1977ur,Peccei:1977hh,BANERJEE2003109}. Even further, there are open
questions about long-standing observations, such as whether or not
there is an extended Higgs sector~\cite{Branco:2011iw}, why there are multiple generations
of fermions with a large mass hierarchy~\cite{Pati:1974yy,1992pmlq.book.....D,Froggatt:1978nt,ArkaniHamed:1999dc},
and why no magnetic monopoles are observed to exist~\cite{doi:10.1063/PT.3.3328}. 
For these reasons, the SM is considered to be an effective field
theory, and that physics beyond the SM (BSM) should exist.

There is no shortage of models to explain these elusive
phenomenon, with varying degrees of complexity and explanatory
power. One very popular group of theories to explain several of these
phenomena involve supersymmetric (SUSY) extensions to the
SM~\cite{Martin:1997ns,Lykken:1996xt}. Many SUSY models contain a particle 
that only interacts very weakly with ordinary matter (the ``lightest
SUSY particle'', or LSP), providing a simple DM candidate. At
the same time, SUSY also attempts to address
questions about the hierarchy problem, the nature of
space-time, grand unified theories, and even string theory. For this
reason, SUSY has long been held as a very attractive BSM physics
model, because it can explain a wide range of
phenomena with simple assumptions.

Unfortunately, as of yet, no easily detectable signals have been
observed at the LHC. This, in and of itself, is not necessarily a
problem, because the scale of SUSY could always either be heavier
than we can currently access, or exists in a region where the signals
are hidden among SM backgrounds. The former case, however, 
limits the ability for SUSY to mitigate the hierarchy problem.The
infrared divergences of the mass of the Higgs boson are only canceled
if the masses of the SUSY particles are very close to their SM
counterparts. This raises questions of whether or not the models
themselves ``naturally'' explain the hierarchy problem. For the case
of subtle signatures, of course, such questions of naturalness are
less pressing, and can still preserve solutions to the hierarchy
problem with a DM candidate. 

Despite those attractive theoretical features,  there is really no a
priori  reason (other than our personal aesthetic) that one model
should address all of these open questions simultaneously. 
For these reasons, in this Review, we will discuss
a subset of
these questions that have been investigated recently at the LHC with
13 TeV proton-proton collisions by the ATLAS, CMS, and
LHCb~\cite{Alves:2008zz} experiments. 
From a collider standpoint, we will discuss the solution to the
hierarchy problem, dark matter, the 
origins of neutrino masses, unification, and compositeness. 
We will also discuss the possibilities for improvements of these
searches at the High-Luminosity LHC (HL-LHC) or other future
colliders.  

With a few exceptions, this Review will focus
on answers to the above questions that do not involve SUSY, although
it remains a theoretically attractive solution. This Review 
will also primarily not focus on solutions that involve an extended
Higgs sector, nor open anomalies in hadron spectroscopy. All of these
topics merit their own separate reviews. 

Many models of BSM physics that can be tested at the LHC often involve
spectacular signatures that distinguish them from SM backgrounds. It
is therefore worthwhile to discuss the searches
for new physics with their unique signatures in mind. As such, we will
first broadly discuss the signatures used for LHC BSM searches, and
then discuss the implications on various scenarios. 

The rest of this Review will be structured as follows. We discuss
novel reconstruction techniques that are used extensively in searches
in Sec.~\ref{sec:tools},
solutions to the hierarchy problem in Sec.~\ref{sec:hierarchy}, 
searches for DM in Sec.~\ref{sec:dm}, 
understanding the neutrino mass in Sec.~\ref{sec:numass}, the
unification of the forces (including leptoquarks) 
in Sec.~\ref{sec:unification}, and finally the compositeness of the
fundamental particles in Sec.~\ref{sec:compositeness}. 
As a guide, Figs.~\ref{fig:exo_atlas}-\ref{fig:exodm_cms} show the
summaries of the searches for non-SUSY BSM physics at ATLAS and CMS
performed with the various techniques outlined in
Sec.~\ref{sec:tools}.

\section{Tools of searches for BSM physics}
\label{sec:tools}

Overall, the major signatures of the
searches for BSM physics will include: (1) traditional signatures
involving 
leptons, jets, and photons with high transverse momentum ($\pt$), or missing transverse momentum ($\PTm$);
(2) signatures involving particles that have lifetimes long enough to
detect their decays (``long-lived particles''); (3) signatures with highly Lorentz-boosted SM
particles that result in collimated, massive jets (``boosted hadronic
jets''); and (4) signatures involving resonances that
decay to lower-mass states, which must be Lorentz-boosted via
initial-state radiation (ISR) to be
detected (``ISR boosted'').

\subsection{Traditional signatures}
\label{sec:tradsig}

The ATLAS and CMS experiments have been designed primarily with
traditional signatures for particle collisions in mind, with
relatively prompt signals containing hadrons and isolated leptons or
photons. The LHCb experiment has slightly different goals, i.e. to
precisely measure bottom and charm hadron production, decays, and
properties, as well as other particles with long lifetimes. 
Of course, 
many models of new physics manifest in SM-like signatures with
different kinematic decays, or at different rates, compared with their SM
counterparts. Considerable effort must occur to ensure optimal
performance of the detectors, triggers, object reconstruction,
calibration, etc. A thorough discussion of the experimental challenges
facing the LHC experiments is beyond the scope of this paper, however
we will highlight a few key ideas that are used in searches for BSM
physics that look qualitatively similar to SM production. 

Hadronic jets are the result of fragmentation and hadronization of the
underlying quarks and gluons in the LHC interactions. Due to the
confinement and asymptotic freedom of the quantum chromodynamic (QCD)
interaction, the fragmentation and hadronization occur primarily in a
collimated spray of particles called
``jets''~\cite{Ellis:2007ib}. They are reconstructed from different
inputs (depending on the detector) using the {\sc \small fastjet}
software package~\cite{Cacciari:2005hq,Cacciari:2011ma}.
The ATLAS collaboration utilizes primarily topological clustering of
their calorimeter deposits (TC)~\cite{Aad:2016upy}, or occasionally a full reconstruction of the
particle flow throughout the detectors
(PF)~\cite{Aaboud:2017aca}, while CMS utilizes PF almost exclusively
except where noted~\cite{Sirunyan:2017ulk}.   
The typical momentum resolutions and scale uncertainties achieved for
both experiments are $\sim 10\%$ and $\sim 0.5$--$1.0\%$, respectively, for 
$\pt = 100$ GeV~\cite{Aaboud:2017jcu,Aaboud:2017aca,Khachatryan:2016kdb}.
Jets containing bottom or charm hadrons can have some displaced
particles within them, and ATLAS, CMS, and LHCb are able to discern very
small displacements (a few tens of microns) with respect to the beam
axis with dedicated tagging
algorithms~\cite{Aaboud:2018xwy,Sirunyan:2017ezt}. This allows the
reconstruction of vertices a few hundred microns from the beam
axis. Such information can be used to efficiently discriminate jets
that originate from bottom or charm quarks from those that originate
from lighter quarks or gluons. 

Electrons and photons are reconstructed in both experiments accounting
for interactions with the material of the detector using dedicated
algorithms~\cite{ATL-PHYS-PUB-2017-022,Aaboud:2018yqu,Khachatryan:2015hwa,Khachatryan:2015iwa},
and using both the electromagnetic calorimeter and tracking information. 
Muons are reconstructed using dedicated detectors outside of the
calorimeter structures~\cite{Aad:2016jkr,Chatrchyan:2012xi},
as well as information about the muon track and the ionization
deposits in the calorimeters. 
The performance is dependent on the purity of the signal in question,
but a good benchmark is the performance in reconstructing electrons
from $Z$ bosons, where the experiments achieve electron momentum resolutions
and scale uncertainties around $1.5$--$5.0\%$ and $<1\%$, respectively,
and muon momentum resolutions and scale uncertainties around $1\%$ and
$1$--$2.0\%$, respectively.

The reconstruction of $\tau$ leptons is performed by first reconstructing jets,
then applying selection criteria consistent with individual
particle signatures that take advantage of the unique decays of the
$\tau$ lepton either hadronically to one or three pions, or
semileptonically to lighter leptons and neutrinos~\cite{Aad:2015unr,Sirunyan:2018pgf}.
There is an additional challenge in $\tau$ reconstruction, in that
there are neutrinos produced in their decay that escape detection, 
which causes difficulties in reconstruction of the four-vector. 
The momentum resolutions and scale uncertainties are around
$15\%$ and  $0.5$--$1.0\%$ for $\tau$ leptons decaying from $Z$ bosons,
respectively.  

Neutrinos are produced at the LHC primarily through weak interactions
of the $W$ boson. They can be produced directly through on-shell $W$
decays, or indirectly via weak decays of bottom or charm quarks, or
$\tau$ leptons. Neutrinos are not directly detected. Their presence is
inferred by taking advantage of the fact that, since the proton beams
carry minimal transverse momentum, the vector sum of the transverse
momenta of all of the observed particles should cancel. This is
referred to as a ``transverse momentum imbalance'' or ``missing
transverse momentum'' $\PTm$. This technique can also be used to signal the
presence of other particles that are not directly detected, such as
DM or other exotic particles.
A critical feature of this method of detection is to have nearly
hermetic coverage of the phase space, but perfect coverage is
unrealistic. This incomplete coverage in part contributes to the
$\PTm$ resolution, which is around $10$--$15\%$ in control samples
involving $Z$ boson decays to $e^+e^-$ and $\mu^+\mu^-$.

\subsection{Long-lived particles}
\label{sec:llsig}

It is possible for some particles that are produced in the collision
to decay after traveling a relatively long distance. The most
colloquially well-known particles in this category are muons and pions, as produced copiously
via interactions of cosmic rays with the upper atmosphere. The
mechanics behind such long decay times can differ, but broadly, there
is either a massive force mediator (such as the $W$ boson) that
weakens the interaction strength, or the masses of the parent and
child particles in the decay are so close that the kinematic phase space for the
decay is restricted. In either of these cases, the probability for the
particle to decay at a given time is reduced, causing a longer
lifetime.  

The LHC detectors were not originally intended
to detect particles
that decay further than a few centimeters from the beamline. The focus
has traditionally been on detecting jets containing bottom or charm
quarks, which decay a few hundred micrometers from the beam axis. Most
other particles are considered to be effectively stable on the
timescales via which they traverse the detectors. For
instance, accounting for their Lorentz boosts, both pions and muons
are long-lived enough to avoid decaying within the detector itself. 

However, considerable progress has also been made to detect particles with
intermediate lifetimes (longer than bottom and charm
hadrons, shorter than pions and muons, from millimeters to
meters). There are several strategies
that can be employed here, and we discuss some of the most popular. 
Firstly, the same strategy as the bottom
and charm hadron detection can be used, whereby particles with long
lifetimes will have large impact parameters with respect to the beam axis. For
instance, in Refs.~\cite{Aaboud:2017iio,Sirunyan:2017jdo}, the
detectors can discern particles that decay tens of millimeters away
from the beam axis. Secondly, signals of events in the calorimeters
that occur outside the beam crossing can be used as in
Ref.~\cite{Sirunyan:2017sbs}. In this case, particles may be produced
with long enough lifetimes to escape the inner detectors, becoming trapped by the nuclear material of
the hadronic calorimeter, to decay some time later. Thirdly, the
particles may be heavy and quasi-stable, leaving large amounts of
ionizing radiation in the tracking detectors.

Newer ideas include proposals of dedicated satellite
experiments outside of the detector collision halls, such as the
``MAssive Timing Hodoscope for Ultra Stable neutraL pArticles''
(MATHUSLA)~\cite{Chou:2016lxi,Curtin:2018mvb} and 
``ForwArd Search ExpeRiment at the LHC''
(FASER)~\cite{Feng:2017uoz} detectors. The former will be able to
detect particles produced in LHC collisions that decay several hundred
meters from the interaction point, which is the same scale as allowed
values
from constraints imposed by Big Bang Nucleosynthesis
(BBN)~\cite{Kawasaki:2004qu,Jedamzik:2006xz}. The latter will be
situated close 
to the beamline, downstream from LHC collisions, to detect long-lived
particles that subsequently decay to lepton pairs. 
Such satellite experiments show
strong promise in extending the reach of discovery of new particles
with long lifetimes.

\subsection{Boosted hadronic jets}
\label{sec:boostedsig}

Particles with masses above the scale of the SM are widely expected in
many BSM scenarios. If these particles have couplings to the heavier SM
particles (and they must, if we are to produce them at the LHC), then
often they contain couplings to top
quarks and $W/Z/H$ bosons. In these cases, due to the large difference
in masses between the BSM particle and the SM particles, the latter
will be produced with large Lorentz boosts. This causes the decay
products of the unstable SM particles to be highly collimated. We
refer to these as ``boosted
objects''~\cite{Abdesselam:2010pt,Altheimer:2012mn,Altheimer:2013yza,Adams:2015hiv,Larkoski:2017jix,Asquith:2018igt}.

In the case of particles that decay fully leptonically such as
$Z\rightarrow \ell^+ \ell^-$, there are some modest adjustments to
identification criteria that distinguish this case from traditional
reconstruction techniques in
Sec.~\ref{sec:tradsig}. These involve nonstandard reconstruction
techniques with relaxed isolation requirements, since
the resulting leptons typically appear geometrically close to other
objects. 

Particles that decay hadronically (such as $H \rightarrow
b\overline{b}$ or $t \rightarrow Wb \rightarrow q\overline{q'}b$) or
semileptonically (such as $t \rightarrow Wb \rightarrow l\nu b$) pose
more of a challenge. The reason is that hadronic particles, as mentioned in
Sec.~\ref{sec:tradsig}, already tend to fragment and hadronize in
regions with small spatial extent. As such, the signatures of boosted
hadronically decaying particles can look quite similar to traditional
jets. Special techniques involving the substructure of jets have been
developed to distinguish boosted hadronically decaying particles from
standard jets. 

Since these techniques are somewhat new, the full phase space of
possibility has not yet been explored for performance
improvements. Some advances can come from better theoretical
understanding of the underlying radiation patterns of jets, and/or
from new advances in machine learning to better distinguish various
types of jets~\cite{Asquith:2018igt}. 

\subsection{ISR-boosted particles}
\label{sec:isrsig}

Oftentimes, particles can be produced that create no detector signature (such as
neutrinos or DM) or signatures that are completely
overwhelmed by SM backgrounds (such as hadronic decays of the W or Z
bosons). Reconstruction of such particles is impossible with standard
techniques at the LHC. 

In order to solve this problem, one clever idea is to look for
signatures that recoil against initial-state radiation
particles such as gluons. With sufficient Lorentz boosts, the
previously undetectable or undiscernible particles become accessible
again. This is the strategy behind most of the searches for DM outlined below, as well as searches for hadronically decaying
BSM particles with masses below the $W/Z/H$ boson masses. This is also
the strategy behind the recent observation of $H\rightarrow
b\overline{b}$~\cite{Aaboud:2018zhk,Sirunyan:2018kst}, and the
observation of hadronic decays of the $W$ and $Z$ bosons while
searching for lower-mass vector resonances in
Ref.~\cite{Sirunyan:2017dnz}.

\section{The hierarchy problem}
\label{sec:hierarchy}

The hierarchy problem is, in its simplest form, a question about
why the electroweak scale (100 GeV) is so much different from the
Planck scale (10$^{18}$ GeV). There are many references that describe
this in detail (for instance,
Refs.~\cite{Martin:1997ns,PhysRevD.98.030001}), so here we discuss
only the broadest overview. 

The Higgs potential can be written as
\begin{equation}
V = m_H^2|H|^2 + \lambda |H|^4.
\end{equation}
where $V$ is the Higgs potential, $H$ is the Higgs field, $m_H$ is the
$\bar{\mathrm{MS}}$ mass of the Higgs boson, and $\lambda$ is a free
parameter, experimentally determined by the vacuum expectation value
(vev). The vev is nonzero if $\lambda > 0$ and $m_H^2 < 0$,
resulting in $\left< H \right> = \sqrt{-m_H^2 / 2\lambda}$,
where $\left< H \right> = 174$ GeV and the observed Higgs mass is
around 125 GeV, yielding $m_H^2 = -(92.9\;\mathrm{GeV})^2$.

\begin{figure}
  \caption{\label{fig:higgsloop} Contributions to the Higgs boson mass
    from quantum mechanical effects. (Taken from Ref.~\cite{Moortgat-Picka:2015yla}).}
  \centering
\includegraphics[width=0.3\textwidth]{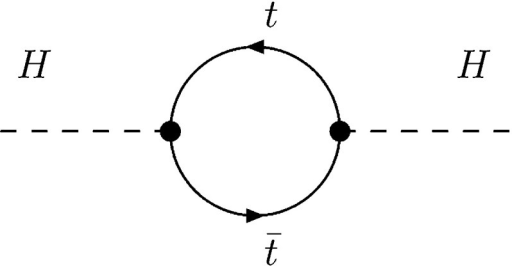}
\end{figure}

The issue arises when one considers couplings of the Higgs field to SM
fermions such as the top quark, in Fig.~\ref{fig:higgsloop}. These
diagrams result in higher-order corrections to $m_H$ such as
\begin{equation}
\Delta m_H^2 = -\frac{ |\lambda_f|^2}{8\pi^2} \Lambda_{UV}^2 + \ldots, 
\end{equation}
where $\lambda_f$ is the Yukawa coupling of the fermion $f$ to the
Higgs field, and $\Lambda_{UV}$ is some upper cutoff of the matrix-element integral
to yield a finite result. There is no physical mechanism within the SM
itself to yield a small value of $\Lambda_{UV}$ to arrive at the
observed Higgs boson mass, so either the SM is valid up to the
Planck scale (resulting in $\Lambda_{UV} = \Lambda_{\mathrm{Planck}}$,
necessitating extremley fine-tuned higher-order corrections,
or a new physical scale exists, $\Lambda_{\mathrm{BSM}}$, between the 
electroweak and Planck scales, interpreted as the scale of BSM
physics. 

There are several proposals for the nature of BSM physics to solve the
hierarchy problem, including SUSY~\cite{Martin:1997ns,Lykken:1996xt},
new strong dynamics or technicolor~\cite{Lane:2002wv,Hill:2002ap},
and extra dimensions, either large~\cite{ArkaniHamed:1998rs,ArkaniHamed:2001nc}
or warped~\cite{Randall:1999ee,Randall:1999vf}.  
Production of signatures involving 
``prompt'' SUSY (i.e., SUSY without long-lived particles) will not be discussed in this
Review, although signatures of SUSY with large lifetimes are discussed
as they overlap significantly with signatures from other models~\cite{Graham:2012th,FARRAR1978575}. 
Large extra dimensions (LED) are discussed below. 
Strong dynamics and warped extra dimensions are linked by an
AdS/CFT correspondence~\cite{Maldacena:1997re}, and are discussed
together using the language of extra dimensions. 

The solutions to the hierarchy problem and unification (see below) 
often predict additional gauge bosons. It is often convenient to
simply assume SM-like couplings in the ``sequential'' SM
(SSM). These are usually taken as benchmark scenarios and
overlap with signatures from other models. 

\subsection{Large extra dimensions}
\label{sec:led}

The existence of large extra dimensions
(LED)~\cite{ArkaniHamed:1998rs,ArkaniHamed:2001nc} solves the
hierarchy problem by positing that gravity is distributed through a
higher-dimensional space (the ``bulk'') whereas the SM particles are
confined to a subspace (the ``SM brane''). This results in a natural
value for $\Lambda_{UV}$, much smaller than 10$^{18}$ GeV. The
relevant parameters are the number of extra dimensions $n$, the
corresponding fundamental Planck scale $M_D$, and the mass threshold
$M_{th}$, above which black holes are formed (where $M_{th} \ge M_D$). The relationship between
$M_D$ and the 3-dimensional Planck mass $M_{pl}$ is given by
\begin{equation}
\label{eq:led1}
M_D = \frac{1}{r}\left( \frac{r M_{pl}}{\sqrt{8\pi}} \right)^{\frac{2}{n+2}}
\end{equation}
where $r$ is the compactification radius.

There are many signatures for LED models, including copious production
of microscopic black
holes~\cite{Banks:1999gd,Dimopoulos:2001hw,Giddings:2001bu}. These black holes
decay almost instantly into one or more particles at high
$\pt$, including signatures with photons, leptons, jets, or
$\PTm$. This provides a very unique signature at the LHC. 
For black hole masses far above $M_{th}$ (for instance, $\sim 4
M_{th}$ for $n=6$)~\cite{Meade:2007sz}, the semiclassical
approximation holds where quantum effects of individual gravitons can
be neglected, and the black hole will decay uniformly to all SM
particles (with quarks and gluons obtaining an enhancement from their
3 colors). The signature of such models contains a large number of high-$\pt$
particles, and so the sensitive variable will be the scalar sum of the
$\pt$ of all of the jets, leptons, photons, and $\PTm$. 
For black hole masses near $M_{th}$, however, the semiclassical
approximation is invalid, and quantum-mechanical decay to a few highly
energetic particles is the dominant decay mode. 

\begin{figure}
  \caption{\label{fig:exo_atlas} Summary of exotica searches at
    ATLAS with traditional, boosted, and ISR-boosted reconstructed
    techniques from Ref.~\cite{ATLAS_EXO_SUMMARY}. These are interpreted in terms of limits on the mass
    of new particles in models containing extra dimensions,
    extra gauge bosons, new contact interactions
    (CI), dark matter (DM), leptoquarks (LQ), heavy quarks, excited
    fermions or miscellaneous others. 
    Yellow (green) bands indicate 13 TeV (8 TeV) data results. }
  \centering
    \includegraphics[width=\textwidth]{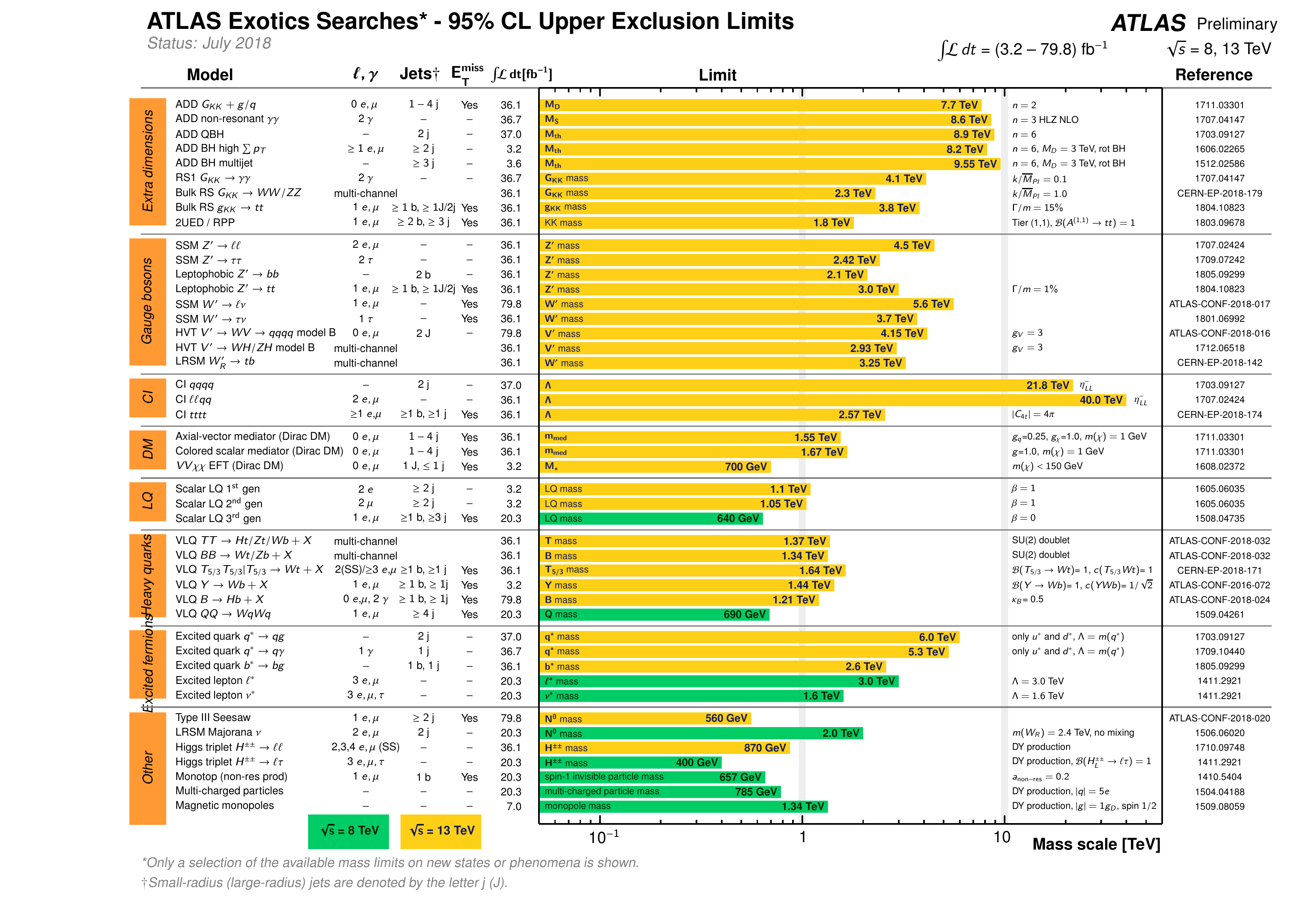}
\end{figure}

\begin{figure}
  \caption{\label{fig:exo_cms} Summary of exotica searches at
    CMS with traditional and ISR-boosted reconstructed
    techniques from Ref.~\cite{CMS_EXO_SUMMARY}. These are presented in terms of limits on the mass of
    new particles in models containing leptoquarks, RS gravitons,
    heavy gauge bosons, excited fermions, multijet resonances, large
    extra dimensions, and compositeness. Boxed (open) bands indicate
    13 TeV (8 TeV) data results. }
  \centering
    \includegraphics[width=\textwidth]{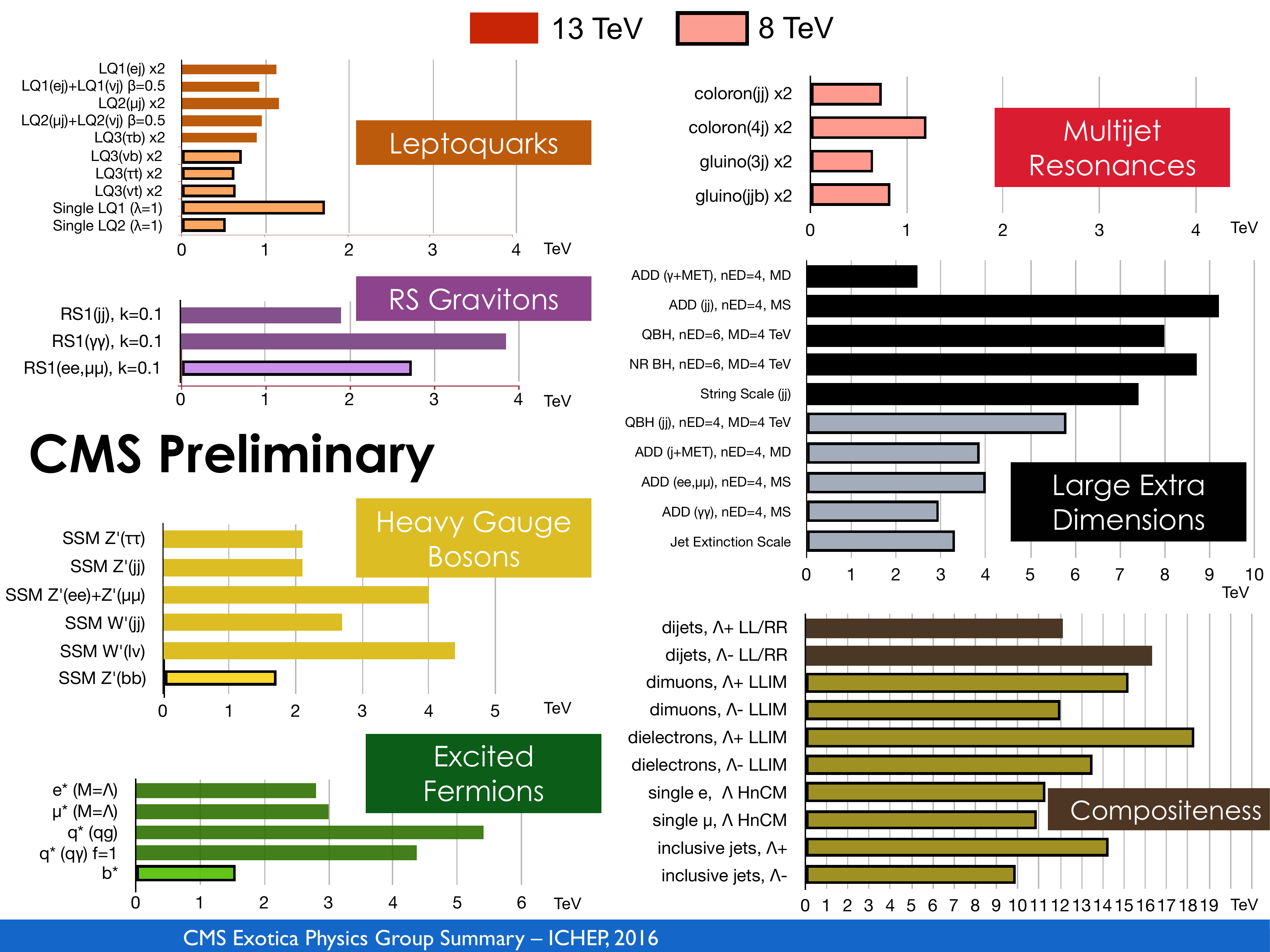}
\end{figure}

\begin{figure}
  \caption{\label{fig:b2g_cms} Summary of exotica searches at
    CMS with boosted reconstructed techniques from Ref.~\cite{CMS_B2G_SUMMARY}. These are interpreted
    in terms of limits on the mass of new particles in models
    containing vector-like quarks, resonances decaying to heavy quarks,
    leptoquarks, excited quarks, and resonances decaying to dibosons.}
  \centering
    \includegraphics[width=\textwidth]{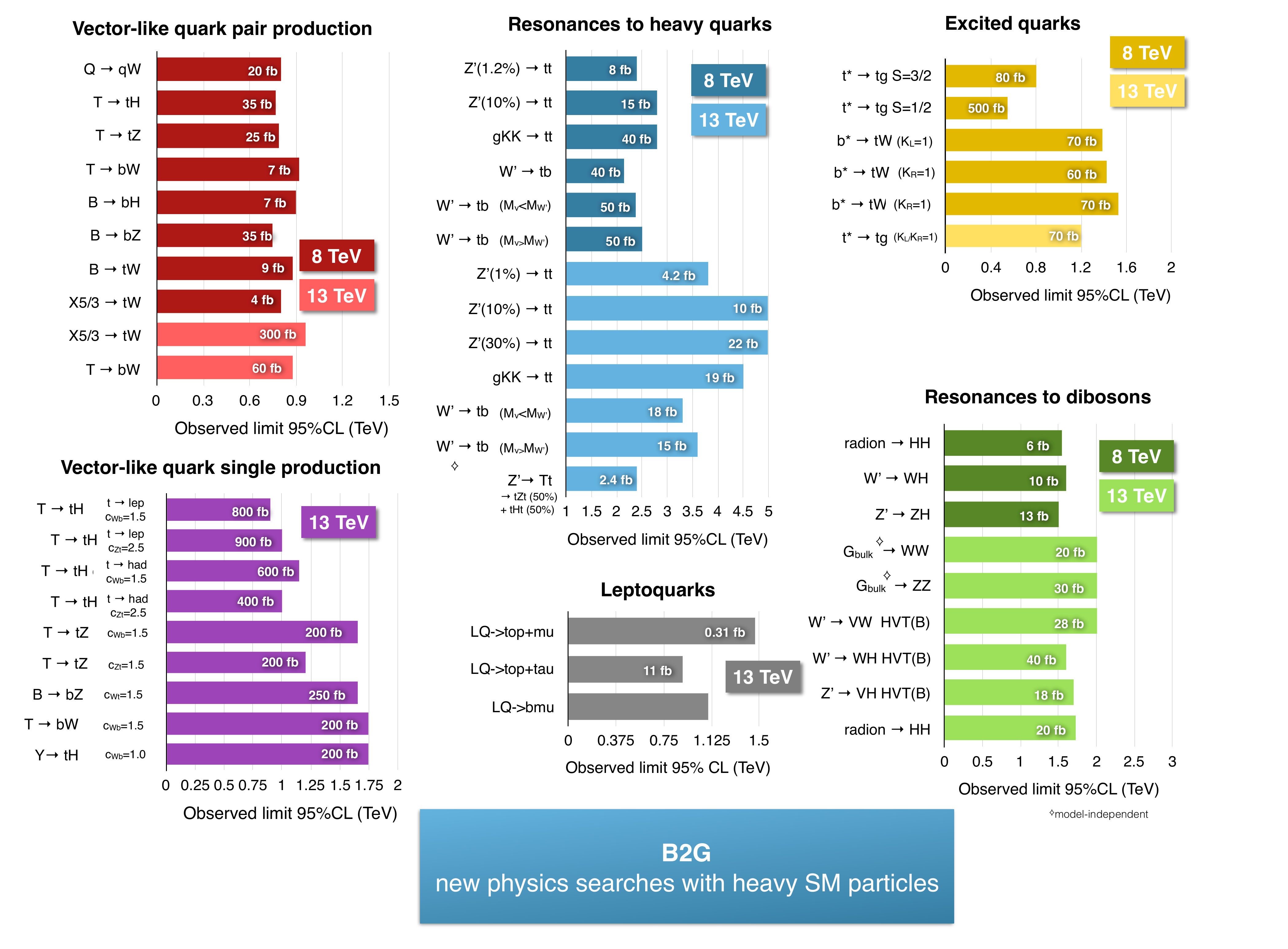}
\end{figure}

At 13 TeV, there have been a large number of searches for such
particles at both
ATLAS~\cite{Aad:2015mzg,Aad:2015ywd,Aaboud:2016uro,Aaboud:2016ewt} and
CMS~\cite{Sirunyan:2017anm,Sirunyan:2017ewk,Sirunyan:2017jix}. 
Figures~\ref{fig:exo_atlas}-\ref{fig:b2g_cms} show the results of many
searches involving high-multiplicity events or events with significant
$\PTm$. The mass limits depend on the signature, the model, and the
number of extra dimensions, but are typically between 2-10 TeV. 
This covers a significant dynamic range of interest for these models
for the case of $n=4$ spatial dimensions, since models with
considerably higher masses would be less likely to solve the hierarchy
problem naturally. 

The energy range of LED models is very large. As such, increases in
the center-of-mass energy will provide the strongest improvements in
sensitivity. However, better estimation of SM backgrounds or improved
analysis techniques can also
lead to improvements with more data at the HL-LHC.  

\subsection{Warped extra dimensions}

Extra-dimensional alternatives to LED include the ``RS1''~\cite{Randall:1999ee} and
``RS2'' models~\cite{Randall:1999vf}. The RS1 model hypothesizes
compact extra dimensions with two branes, one at the Planck scale
and the other at the TeV scale. The SM particles are presumed to exist
primarily on the TeV brane, and have Kaluza-Klein (KK) excitations
around the TeV scale, which behave similarly to their SM counterparts
and hence can be detected at colliders like the LHC. The RS2 model is
similar to RS1, but omits the brane at the TeV scale, and also yield
a KK tower of particles corresponding to the existing SM particles. 

RS1 models can produce black holes as in Sec.~\ref{sec:led}, with
masses larger than
\begin{equation}
M_D = \frac{M_{pl}}{\sqrt{8\pi}} e^{-\pi kr}
\end{equation}
where $r$ and $M_{pl}$ are defined in Eq.~\ref{eq:led1}, and $k$ is a
warp factor. These models also result in KK excitations of the graviton~\cite{Fitzpatrick:2007qr} and
gluon~\cite{Agashe:2006hk,Lillie:2007yh}, which can yield signatures
in many final states such as dibosons, diquarks, di-Higgs, diphotons,
and many others. One common feature is the high masses of the KK
excitations, which often subsequently decay to highly Lorentz-boosted
SM particles, necessitating the usage of the techniques outlined in
Sec.~\ref{sec:boostedsig}. Such models also can result in
additional quarks and/or leptons that transform as vectors under the
ordinary symmetry of the SM, referred to as ``vector-like''
quarks (VLQs) or leptons (VLLs) ~\cite{DeSimone:2012fs}.

Typically, the simplest signatures involving RS models (or the SSM) are resonances
that decay to two objects. There are dilepton~\cite{Khachatryan:2016qkc,Aaboud:2018bun,Sirunyan:2018exx,Khachatryan:2016jww,Aaboud:2016zkn,Aaboud:2017efa,Aaboud:2018vgh},
diphoton~\cite{Khachatryan:2016hje,Khachatryan:2016yec,Aaboud:2016tru,Aaboud:2017yyg,Sirunyan:2018wnk}, 
jet+boson or
diboson~\cite{Sirunyan:2016wqt,Khachatryan:2016odk,Sirunyan:2017hsb,Sirunyan:2017jix,Sirunyan:2017uhk,Aaboud:2018bun,Aaboud:2018fgi,Aaboud:2018eoy,Sirunyan:2018hsl,Aaboud:2017cxo,Aaboud:2017fgj,Aaboud:2017itg,Aaboud:2017eta,Aaboud:2017ahz,Aaboud:2017dor,Aaboud:2016trl,Aaboud:2016lwx,Aaboud:2016okv,Sirunyan:2017hsb,Sirunyan:2018iff,Sirunyan:2017yrk,Sirunyan:2017isc,Sirunyan:2017acf,Sirunyan:2017wto,Sirunyan:2017nrt,Sirunyan:2016cao,Aaboud:2016xco}, and
diquark/dijet~\cite{Khachatryan:2015dcf,Aaboud:2017yvp,Aaboud:2018fzt,Khachatryan:2015dcf,Sirunyan:2016iap,Sirunyan:2017ygf,Sirunyan:2018wcm,Sirunyan:2018xlo} analyses. 
There are also specialized diquark/dijet analyses in resonant
production of
$b\overline{b}$~\cite{Aaboud:2016nbq,Aaboud:2018tqo},
$t\overline{t}$~\cite{Sirunyan:2017uhk,Aaboud:2018mjh,Sirunyan:2018ryr}, 
$t\overline{b}$~\cite{Aaboud:2018juj,Sirunyan:2017ukk,Sirunyan:2017vkm}, 
and
resonances decaying to VLQs~\cite{Sirunyan:2017bfa}. 
Overall, the benchmarks used in these searches are RS1 KK gravitons, RS1 KK
gluons (for $t\overline{t}$ resonances), or $W'$ bosons (for
$t\overline{b}$ resonances). 
There are also other models that are probed with the dijet and
$b\overline{b}$ resonance papers. 
The
limits on these models are already quite stringent, effectively
saturating the available parton luminosity at high masses in the multiple TeV range.
There are also analyses that manifest as a combination of ISR boosts
as in Sec.~\ref{sec:isrsig} and boosted hadronic jets as in
Sec.~\ref{sec:boostedsig}, shown in Ref.~\cite{Sirunyan:2017dnz}
There are also many analyses searching for direct production of 
VLQs~\cite{Aaboud:2018xuw,Aaboud:2017qpr,Sirunyan:2018omb,Sirunyan:2018fjh,Sirunyan:2018ryt,Sirunyan:2017ynj,Sirunyan:2017usq,Khachatryan:2016jww,Sirunyan:2017tfc,Sirunyan:2017ezy,Sirunyan:2016ipo,Khachatryan:2016vph}.

Updates to these analyses will need to predominantly start focusing on
reducing the SM background and its uncertainty, until a new collider
is built at a significantly higher energy. In many cases, the
resonances at higher masses are so  broad that they are predominantly
produced away from the resonant peak (``off-shell''),  and manifest
like a contact interaction above the SM backgrounds.  In the case of a
signal at lower  mass, it will be difficult   to interpret the precise
mass of the new physics signals because of this off-shell effect.
There is still sensitivity in    the lower-mass states with increasing
luminosity, so the HL-LHC will continue to provide useful improvements
in these searches.

\section{Dark matter}
\label{sec:dm}

Dark matter comprises 4-5 times as much of the universe as ordinary
matter~\cite{Ade:2013sjv}. It is natural to suppose that DM is comprised of
particles that interact very seldomly, i.e. that it is due to ``weakly
interacting massive particles,'' or WIMP. The relic density of DM hints at particle DM at the electroweak scale (the
``WIMP miracle'')~\cite{Bertone:2004pz,Feng:2010gw,Porter:2011nv}. 
However, as of now, we have no candidate particle to explain the
evidence. This remains one of the major open questions in physics.

As
mentioned above, this review will not discuss the overall state of the
search for SUSY, leaving this to other reviews, but instead we will
focus on specific SUSY-inspired final states that include signatures
that are difficult to detect (``hidden''). 

While SUSY does provide a single natural DM candidate, there
is nothing constraining the particle content of the dark sector. There
may be a family of dark particles, even with 
their own interactions, that comprise the dark sector. The only real
constraint we have is that if WIMPs exist, they interact weakly with
SM particles. For this reason, more model-agnostic searches have
become popular, with the help of effective field theories (EFTs) or
simplified models of
DM interactions~\cite{Abercrombie:2015wmb}. These focus
more on the signatures involving DM and place constraints
simultaneously on the masses of the DM, and the mediator via
which they interact with the SM particles. An exhaustive list of final
states with spin hypotheses of the mediator can thus be made, and an
extensive program has been undertaken to investigate these models. 

We will now investigate the phenomenology of hidden signatures, as
well as that of EFTs/simplified models in detail.

\subsection{Hidden sectors and RPV SUSY}
\label{sec:hierarchy_hidden}

The postulation of a hidden
sector~\cite{Barbieri:2005ri,Schabinger:2005ei,Strassler:2006im} can explain
DM, and arises in many solutions to the hierarchy
problem. Some models postulate a non-abelian sector of light particles
that interacts with the SM via a
heavy mediator, thus becoming ``hidden'' or ``dark''. These particles
could form complex bound states since they are strongly interacting,
thus forming ``valley hadrons'' or ``v-hadrons'' analogous to QCD. The
LHC could in principle produce these v-hadrons, which would
subsequently decay to detectable SM particles through the massive
mediators after a long time~\cite{Strassler:2006ri}, resulting in
observable SM particles that are displaced from the interaction point,
analogous to a charged pion that decays to a muon and neutrino via a
massive $W$ boson. This necessitates utilizing the detection
techniques outlined in Sec.~\ref{sec:llsig}. Furthermore, the decay
products may also potentially be collimated, necessitating the
techniques outlined in Sec.~\ref{sec:boostedsig}. 
The Higgs boson could in principle couple with the hidden sector,
providing a ``Higgs portal''~\cite{Englert:2011yb}. 
The latter signature would be a Higgs boson produced and decaying into
long-lived v-hadrons, which may or may not decay to SM particles
within the detector acceptance. 

In addition to model-agnostic hidden sectors, SUSY can result in
signatures that are quite similar, if they violate
R-parity~\cite{Graham:2012th,FARRAR1978575}, i.e. RPV SUSY. In these
cases, the LSP will often be sufficiently long-lived to
decay centimeters or meters away from the LHC collisions. 
The methodologies for detection can range from detection of
particles that decay within the tracker
volume, possibly with other distinguishing features like
$\PTm$~\cite{Aaij:2016xmb,Sirunyan:2017jdo,Aaboud:2017iio},
those that contain extensive ionizing radiation in the
tracker~\cite{Khachatryan:2016sfv}, particles that decay
into hadronizing particles far from the interaction region
(``emerging'' jets)~\cite{Aaij:2017mic}, particles that get trapped in the nuclear
material and subsequently decay~\cite{Sirunyan:2017sbs},
particles that decay to unobservable particles in flight
(``disappearing'' tracks'')~\cite{Aaboud:2017mpt}, and others not
discussed here.

\begin{figure}
  \caption{\label{fig:exoll_atlas} Summary of long-lived exotica
    searches at ATLAS from Ref.~\cite{ATLAS_LL_SUMMARY}. These are
    interpreted in terms of limits on the lifetime of various models. }
  \centering
    \includegraphics[width=\textwidth]{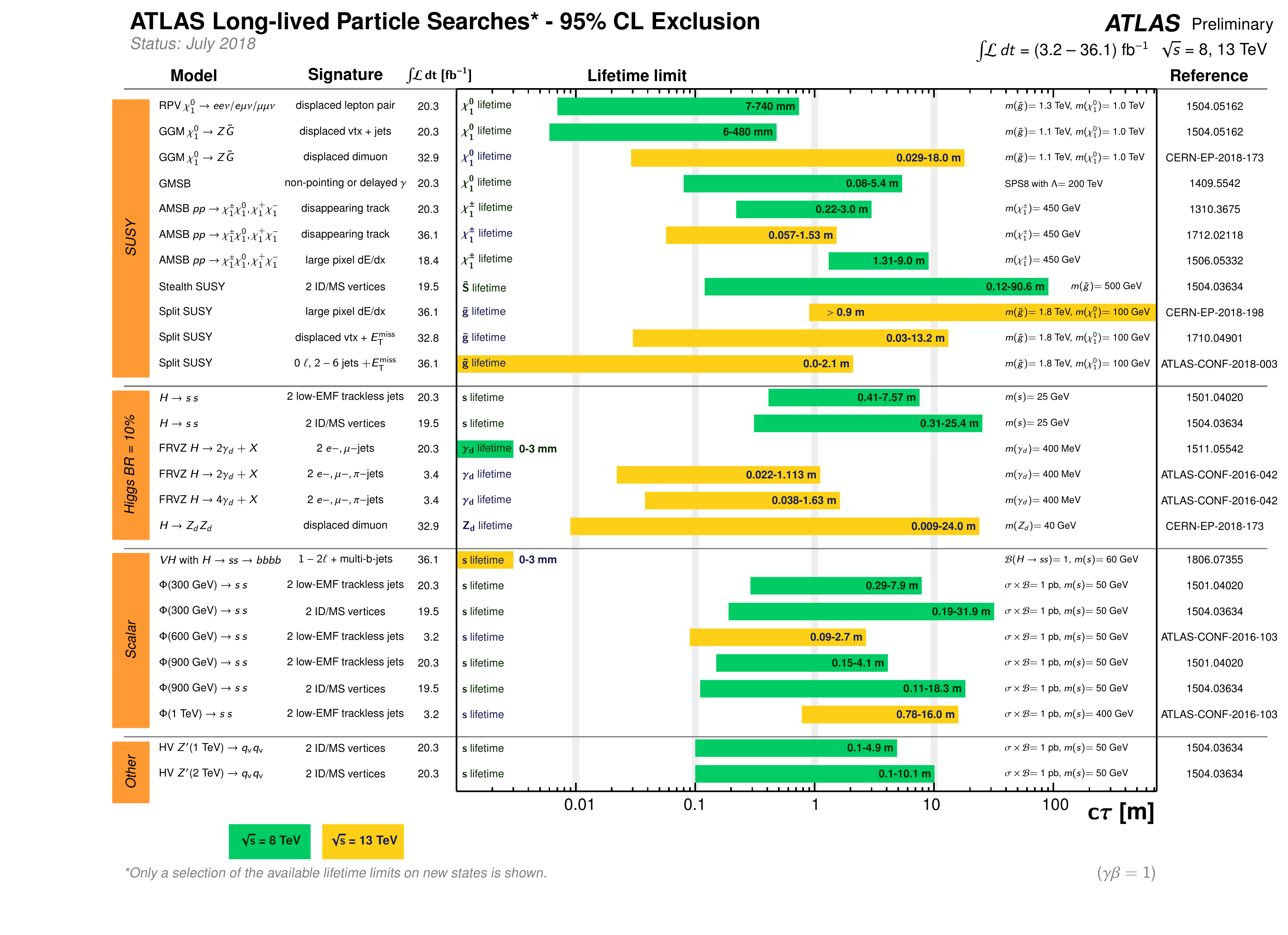}
\end{figure}

\begin{figure}
  \caption{\label{fig:exoll_cms} Summary of long-lived exotica
    searches at CMS from Ref.~\cite{CMS_LL_SUMMARY}. These are 
    interpreted in terms of limits on the lifetime of various models.}
  \centering
    \includegraphics[width=\textwidth]{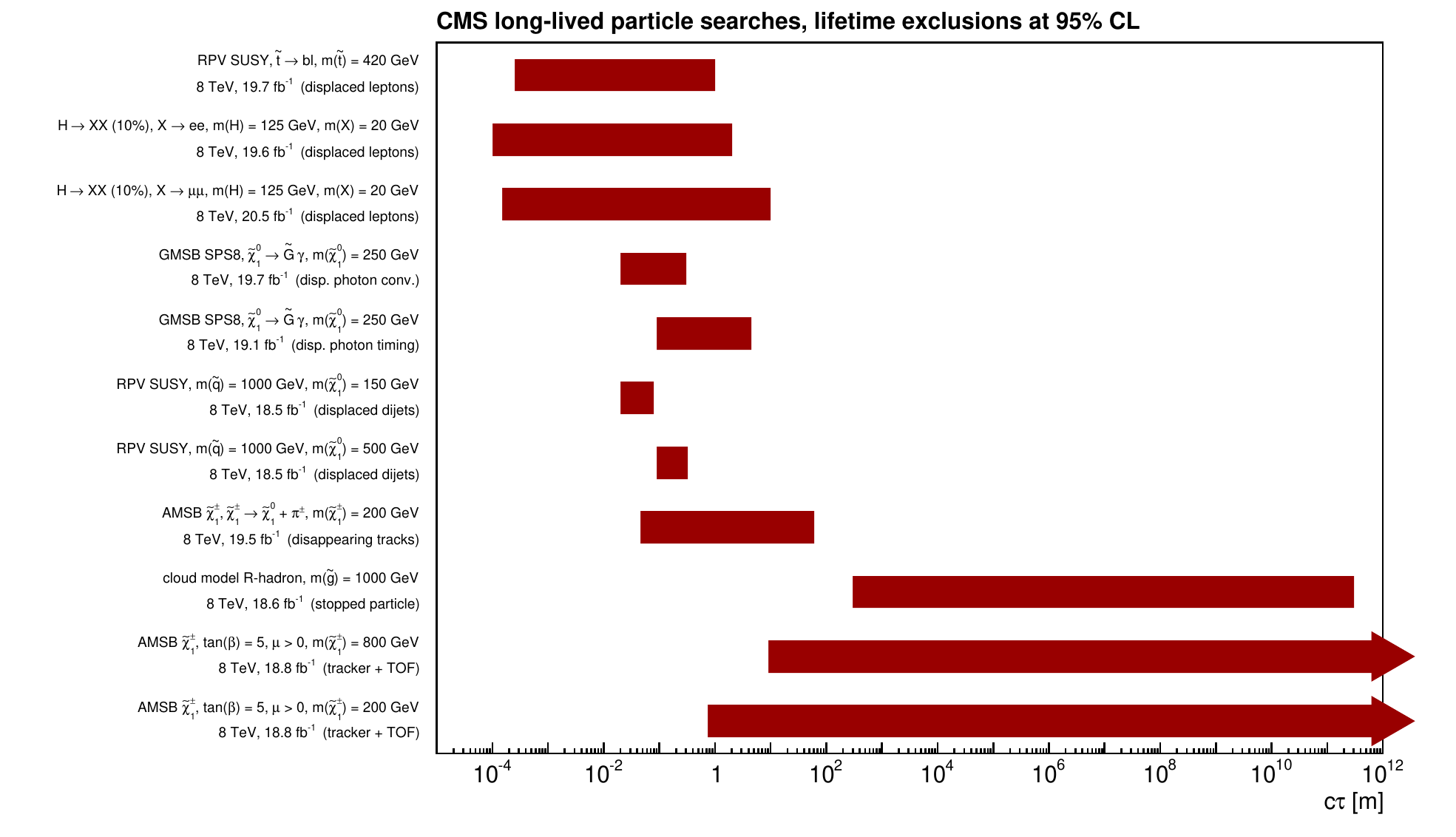}
\end{figure}

Figures~\ref{fig:exoll_atlas} and~\ref{fig:exoll_cms} show summary
plots from ATLAS and CMS of searches for long-lived signatures from
various models. An impressive array of models has been investigated
at a wide range of distances over 15 orders of magnitude, ranging from
millimeters to many meters at very long times. 

Future directions of these searches will predominantly involve
extending the baseline of detection or searches. Projects such as
MATHUSLA and FASER are extremely promising ways to extend the reach and
capability of these types of searches. It is still quite possible that
natural SUSY models (RPV or not) could be found in these difficult
signatures, and it should be a major part of the HEP program in the
future.

\subsection{EFTs and simplified models of DM}
\label{sec:dmsm}

The overall construction of an EFT involving DM postulates a very
massive mediator of the interaction between DM and SM particles, and hence
can be modeled as a contact interaction. 
Simplified models, on the other hand, postulate various
DM--SM mediators, as well as a DM particle, all with varying
spins and couplings to the SM particles. Broadly speaking, these can
both result in similar signatures. Overall, since any DM particles
that are produced in LHC collisions will not interact with the
detectors at all, detection techniques focus primarily on ISR-boosted
detection techniques as in Sec.~\ref{sec:isrsig}, and reconstruct the
observable interaction from ISR with 
traditional techniques as in Sec.~\ref{sec:tradsig} or with boosted
hadronic jets as in Sec.~\ref{sec:boostedsig}. Depending on the final state,
flavor tagging techniques to detect bottom or top quarks can also be
used. As such, existing analyses include a dizzying array of final
states~\cite{Aaboud:2016qgg,Aaboud:2016obm,Aaboud:2017dor,Aaboud:2017uak,Aaboud:2017yqz,Aaboud:2017bja,Aaboud:2017rzf,Aaboud:2017phn,Sirunyan:2016iap,Sirunyan:2017onm,Sirunyan:2017hci,Sirunyan:2017hnk,Sirunyan:2017xgm,Sirunyan:2018gka,Sirunyan:2018wcm,Sirunyan:2018xlo}.
These are usually colloquially referred to as ``mono-X'' searches,
since the signature in the detector is a single particle (X) recoiling
against the DM particle. The particle X can be any SM particle. There
are therefore searches with signatures of
mono-jet, mono-bottom-jet, mono-top-jet, mono-photon, mono-$W$,
mono-$Z$, mono-Higgs, etc. The mediators can also interact with a
pair of particles, so signatures can also involve $q\overline{q}$, $\ell^+\ell^-$, $b\overline{b}$,
$t\overline{t}$, etc. 

Various interaction hypotheses are investigated for the DM--SM
mediators. They can be vectors, axial-vectors, scalars, or
pseudoscalars. The coupling constants for the DM--SM interaction are
also unconstrained, so results must be framed in terms of these
parameters. For instance, Ref.~\cite{Sirunyan:2017jix} present limits on the masses
of a vector mediator and DM (with couplings to SM quarks equal to 0.25) of 1.8
and 0.7 TeV, respectively, in signatures
containing Lorentz-boosted mono-$V \rightarrow qq$. Another example is
Ref.~\cite{Aaboud:2017phn}, which presents limits on the masses of
an axial-vector mediator and DM of 1.5 and 0.4 TeV, respectively, using
a mono-jet signature. 

\begin{figure}
  \caption{\label{fig:exodm_atlas} Summary of searches for DM from
    multijet final states with an axial-vector mediator at ATLAS from
    Ref.~\cite{ATLAS_DM_SUMMARY}. These are interpreted in terms of
    limits on the masses of the mediator and dark matter candidate.}
  \centering
    \includegraphics[width=\textwidth]{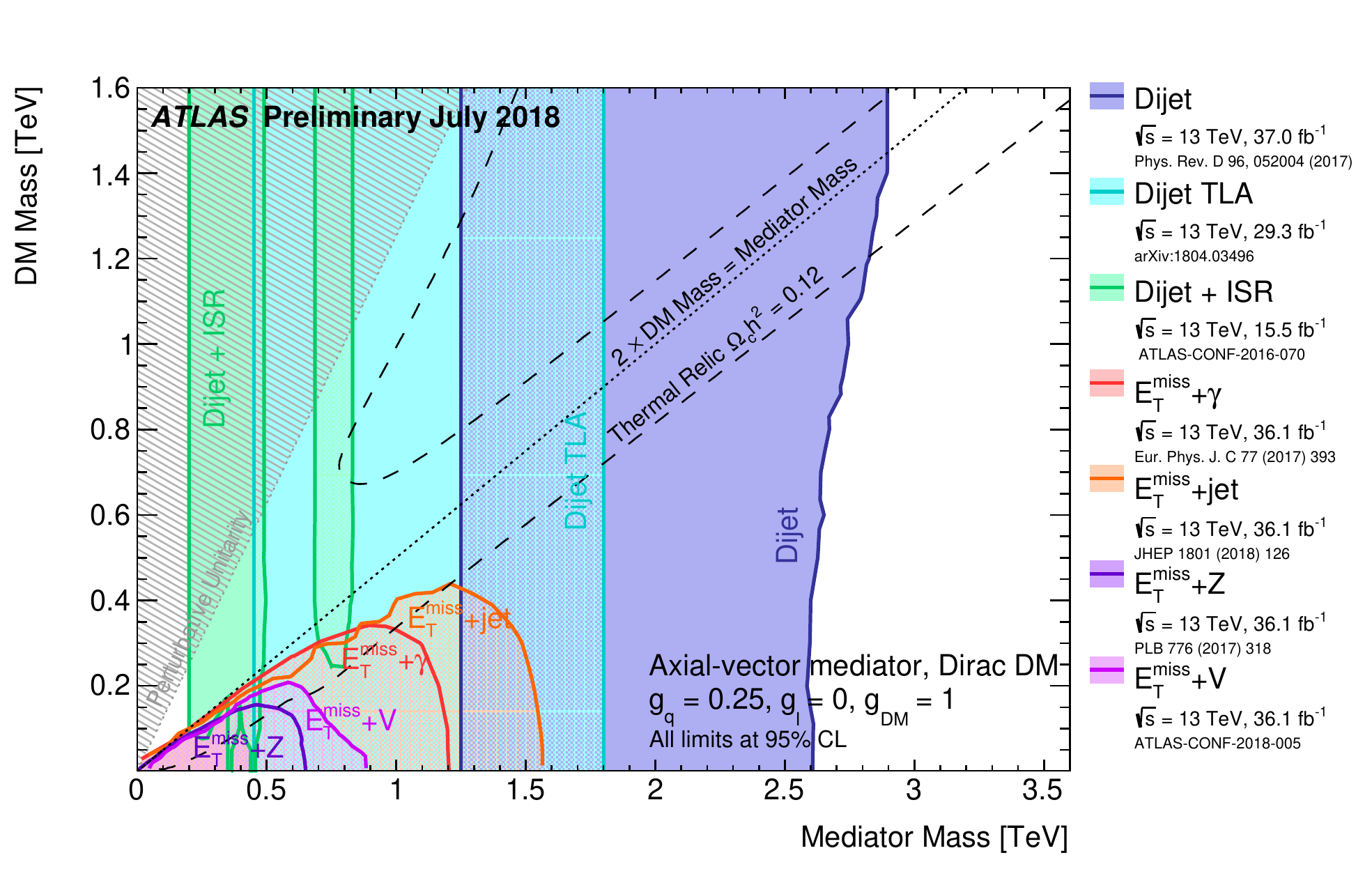}
\end{figure}

\begin{figure}
  \caption{\label{fig:exodm_cms} Summary of searches for DM from
    multijet final states with an axial-vector mediator at CMS from
    Ref.~\cite{CMS_DM_SUMMARY}. These are interpreted in terms of
    limits on the masses of the mediator and dark matter candidate.}
  \centering
    \includegraphics[width=\textwidth]{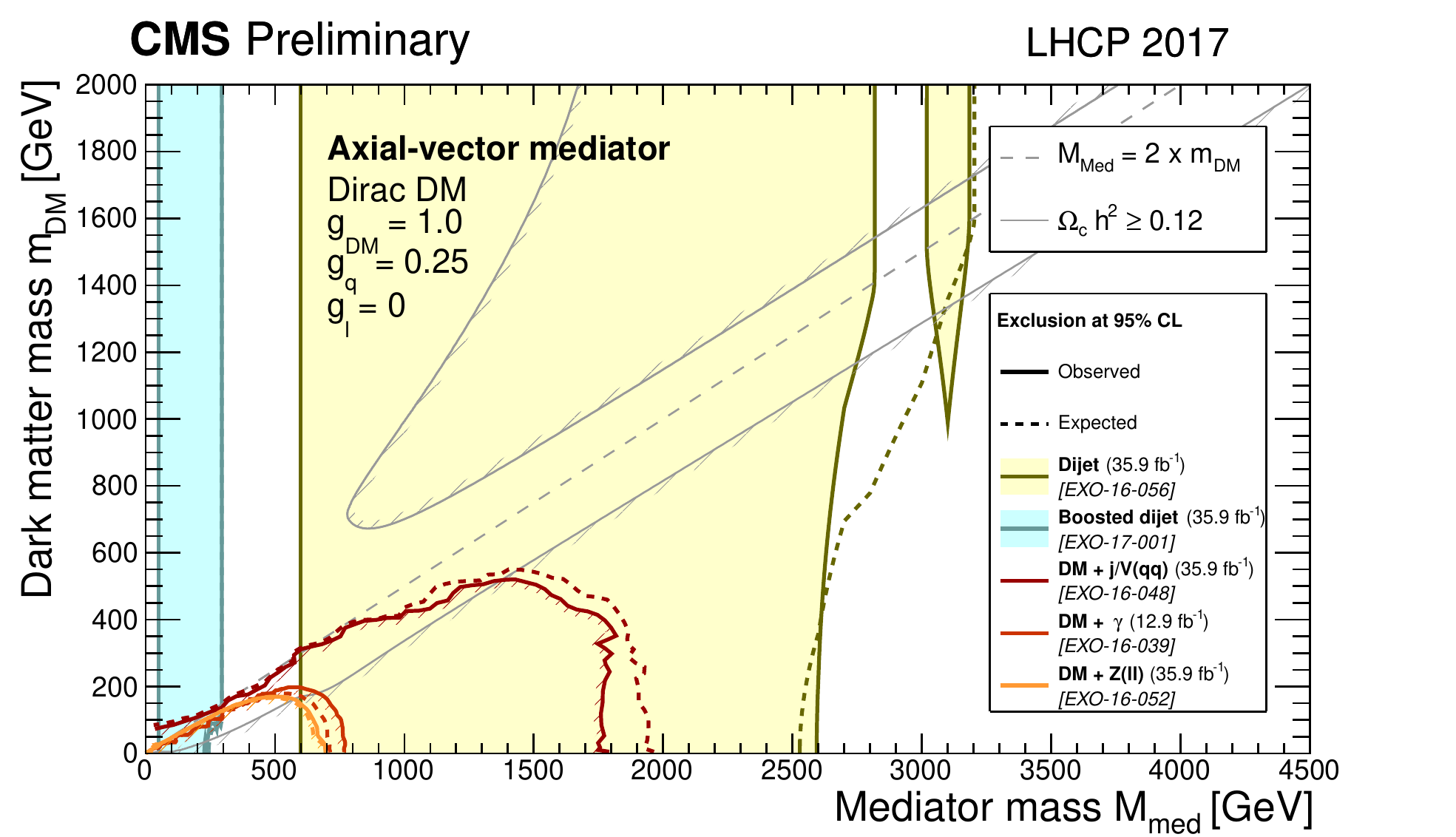}
\end{figure}

In simple interpretations of the DM--nucleon scattering cross section
as a function of the DM mass, LHC searches complement direct
detection (DD) and indirect detection (ID) searches~\cite{Kahlhoefer:2017dnp}. Overall, LHC searches are more
sensitive than ID/DD at very low mediator masses (below 5 GeV), as well as for axial-vector mediators,
whereas ID/DD searches are more sensitive at higher masses if there are
vector or scalar mediators. For instance, for a vector mediator,
Refs.~\cite{Sirunyan:2017jix,Aaboud:2017phn} show DM--nucleon cross-section limits
of $\sim 10^{-42}$ cm$^2$ for a DM mass of 1 GeV, whereas there is no
corresponding DD sensitivity, but the DD searches become more
sensitive for DM masses around 30 GeV, with cross-section limits of
$\sim 10^{-46}$ cm$^2$ from XENON1T~\cite{Aprile:2018dbl}. 
Figures~\ref{fig:exodm_atlas} and~\ref{fig:exodm_cms} show limits of
searches for axial-vector-mediated DM in multijet final states from
ATLAS and CMS, respectively. 

For much of the phase space, the limits can be improved with increased
luminosity. As such, future prospects for DM detection are quite
strong at the HL-LHC. 

\section{Neutrino mass}
\label{sec:numass}

As of yet, the observation of non-zero neutrino masses is the
strongest direct evidence for BSM particle physics. DM also
strongly points to a new sector, but has not been directly observed
nor produced in particle-particle interactions, and the effects are
only observed at large distances, either in galaxial rotations or CMB
observations. Neutrinos, on the other hand, have been directly shown
to have individual masses, and an extensive research program exists to
investigate this regime~\cite{Tortola:2012te}. 

The LHC can play a role in the investigation of such anomalies by
searching for possible heavy partners of the neutrino $N$, which are
naturally predicted by the ``seesaw'' mechanism~\cite{Minkowski:1977sc,Mohapatra:1979ia,Mohapatra:1986bd,An:2011uq}, where the neutrino
masses $m_\nu$ are proportional to $y_\nu^2v^2/m_N$, where $v$ is the
vacuum expectation value of the Higgs field, and $y_\nu$ is a Yukawa
coupling. Very small neutrino masses $m_\nu$ could correspond to large
masses for the heavy neutrinos. It is quite
reasonable to expect that, should such a mechanism exist, the LHC
would be able to observe these partners. There are, as such, many
searches for BSM physics involving heavy neutrinos decaying into
various final states, including leptons, 
jets, or bosons~\cite{Sirunyan:2018omb,Sirunyan:2017xnz,Sirunyan:2017yrk,Khachatryan:2016jqo,Sirunyan:2018mtv}. 

Overall, the exclusion depends on the relative mixing between the
light and heavy neutrinos, $V_{\nu N}$. If this mixing is 0.1,
the masses probed by existing searches are in the several hundred GeV
range. If the mixing is 1, the masses probed are close to 1 TeV. 

Production of heavy neutrinos is mostly limited by the available
center-of-mass energy, so future colliders will be very effective at
extending the reach of searches for heavy neutrinos. There will be,
however, still available phase space to explore at the HL-LHC for
lower masses. 

\section{Unification}
\label{sec:unification}

Extensions to new gauge sectors that encompass the SM have long sought
to find an overarching symmetry that couples the strong and
electroweak forces. 
Fundamentally, any unification of the strong and electroweak forces
will involve some BSM coupling between quarks and leptons. One can
think of this as lepton number being a fourth color. Oftentimes,
such an interaction will contain new particles that contain quantum
numbers for both the strong and electroweak forces. These are known as
``leptoquarks'' (LQs)~\cite{Pati:1973uk,Pati:1974yy,Georgi:1974sy,Murayama:1991ah,Fritzsch:1974nn,Senjanovic:1982ex,Frampton:1989fu,Frampton:1990hz,Schrempp:1984nj,Dimopoulos:1979es,Dimopoulos:1979sp,Buchmuller:1986zs,Eichten:1979ah,Hewett:1988xc}. Of course, such interactions
would also contain predictions for unstable
protons~\cite{Nath:2006ut}, where extremely stringent limits must be
considered in building BSM physics models. 

There is further recent interest in LQs because they have been
proposed as solutions~\cite{Wehle:2016yoi,Altmannshofer:2014rta,Descotes-Genon:2015uva,Hurth:2016fbr,Assad:2017iib,Altmannshofer:2017fio,DAmico:2017mtc,Altmannshofer:2017yso,Calibbi:2017qbu} to several outstanding hints of lepton flavor
non-universality in heavy-flavor
hadron observations from Belle~\cite{Abashian:2000cg} and
LHCb~\cite{Aaij:2016flj,Aaij:2015esa,Aaij:2015oid}. 
Such particles have also been
hypothesized~\cite{ColuccioLeskow:2016dox,Chen:2017hir} to explain the $g-2$ anomaly~\cite{Holzbauer:2016cnd,Bennett:2006fi}. 

With those considerations in mind, many models of unification testable
at the LHC will contain
LQs. Broadly speaking, these will occur as an excess of events
involving both leptons and hadrons. There are various strategies to
deal with such
signatures~\cite{Aaboud:2016qeg,Sirunyan:2018jdk,Sirunyan:2018ryt,Sirunyan:2017yrk,Khachatryan:2016jqo,Sirunyan:2018nkj}.  
One example is to search for first- or second-generation LQs
coupling to first- or second-generation quarks and leptons. In those
cases, analyses can estimate the background for such searches using the known rates of
electroweak production of $W$ and $Z$ bosons, as well as top quark
pair production. Another strategy is to search
for third-generation LQs in signatures involving $\tau$
leptons, bottom or top quarks. The SM backgrounds for such signatures
are dominated by top quark pair production, which can be
predicted. The limits for LQs are currently on the order of 800-1500
GeV depending on the channel.

Since the masses of the LQs the LHC is sensitive to are relatively modest, increases in
luminosity at the HL-LHC can provide a good opportunity to continue
these searches. 

\section{Compositeness}
\label{sec:compositeness}

Ever since Rutherford began to probe the structure of the proton, the
question of whether or not the particles we observe are fundamental or
composite is a perennial question. Investigations of quark
compositeness are not fundamentally different than the Rutherford
experiment, and involve investigations of the number of high-mass
quark-quark interactions. Since a massive mediator would often
manifest as a contact interaction at lower energies (much like the $W$
boson appears as a contact interaction in pion decay, etc), the
searches often focus on such interactions. 
At its heart, the LHC is a QCD jet factory. As such, it can set
extraordinarily sensitive limits on such fundamental
interactions. The searches in Refs.~\cite{Aaboud:2017yvp,Sirunyan:2018wcm,ATLAS:2015nsi}, for
instance, are able to set limits on composite scales between 10-20 TeV. 
The size of the quark is pointlike down to 10$^{-18}$ m, and the scale
of contact interactions manifesting in dijet samples must be
larger than the scale of the LHC center-of-mass energy.

There are also searches for signals of compositeness that search for
excited states of fermions, which then radiate either photons or
gluons with specific characteristics. For example, excited quarks are
investigated in Refs.~\cite{Aaboud:2017yvp,Sirunyan:2017fho,Aaboud:2017nak,Aaboud:2018tqo,Sirunyan:2018xlo}, and dedicated
searches for excited top quarks are shown in
Ref.~\cite{Sirunyan:2017yta}. Excited top quarks are excluded below 1
TeV, and excited light quarks are excluded below 3-5 TeV. 

Generally speaking, compositeness is probed by increases in
center-of-mass energy more than by collecting more data. As such, the
HL-LHC prospects for such searches for BSM physics are somewhat
limited. New colliders at a higher center-of-mass energy would
drastically increase the sensitivity. 

\section{Discussion}
\label{sec:discussion}

As of yet, there are no substantive signals of BSM physics at the
LHC. However, it is unwise to conclude that none exist. There is, a
priori, no particularly better region of phase space aside from
arguments about how much tuning we are psychologically comfortable
with in nature. It is indeed true that a great portion of the
available kinematic phase space of the LHC has been ruled out for
strongly produced BSM signatures (with picobarn-level cross sections),
but the new particles may simply
have larger masses than we have excluded at the LHC (i.e. are heavy), 
may have cross sections that are below our current sensitivity, 
decay outside our detector volume, or we have not looked explicitly in the
correct signatures
(i.e. are hidden). There are multiple strategies to deal with
increasing sensitivity to these signatures, based on new detection and
reconstruction techniques.

Of course, for heavy signatures, there is
nothing better than building a new proton-proton collider at a much
higher center-of-mass energy. However, better reconstruction and
background rejection techniques can improve sensitivity
considerably. 
In addition, there are a plethora of
targeted signatures that are not difficult to investigate, but the LHC 
experiments have simply not addressed them. 

Hidden signatures require several approaches. If a particle is
strongly produced, but decays outside of the region where our
traditional techniques are efficient, new strategies must be employed
to be sensitive to them. This includes detection of long-lived
particles via extensions to the CMS and ATLAS detectors such as
MATHUSLA and FASER. Alternatively, there may be direct signatures that are produced with
smaller cross sections than we are currently sensitive to. Such
searches will improve with more accumulated luminosity at the
HL-LHC. These are typically extremely time-consuming searches, because
they require extensive understanding of the background and subtle
systematic effects. A long, arduous program of measurements and signal
characterization is necessary to investigate these BSM signals. Such
signatures could also be produced indirectly via interactions with the
electroweak bosons, or the Higgs. In this case, such signatures will
have much lower cross sections, and again require rigorous
understanding of the SM background.

Overall, the LHC search program has an extensive future in the HL-LHC
era and beyond. We should not give up hope only because our preferred
ideas do not correspond to what actually exists in the universe.

\section{Acknowledgements}

This work has been supported under NSF grant 1806573, ``High Energy
Physics Research at the CMS Experiment''. SR would also like to thank
Patrick Meade for helpful discussions about black hole semiclassical
approximations. 

\section*{References}

\bibliography{exotica_review_2018}

\end{document}